\documentclass[cameraready]{Interspeech}

\usepackage{amsmath,amssymb,booktabs,multirow,bm}
\usepackage{graphicx}

\title{Revisiting Label-Free Speaker Embedding Enhancement with vMF Profile Likelihood}

\author[affiliation={1}, orcid=0009-0004-3373-5991]{Seunghwan}{Kim}
\author[affiliation={1}]{Jinyong}{Kim}
\author[affiliation={1}]{Sooyoung}{Yang}
\author[affiliation={1}]{Youngjin}{Ko}
\author[affiliation={1,2,3}, correspondingauthor]{Myungjoo}{Kang}

\address{
    $^1$ Interdisciplinary Program in Artificial Intelligence, Seoul National University, South Korea \\
    $^2$ Department of Mathematical Sciences, Seoul National University, South Korea \\
    $^3$ Research Institute of Mathematics, South Korea
}

\email{\{seunghwan.kim, jinyong.kim, jimmy1016, yoko2606, mkang\}@snu.ac.kr}


\newcommand{\eps}{10^{-6}}

\makeatletter
\newcommand{\ps@publication}{%
  \ps@empty
  \setlength{\footskip}{18pt}%
  \def\@oddfoot{%
    \normalfont\fontsize{8}{10}\selectfont\color{black!65}%
    \textit{Proc. Interspeech 2026}, pp.~393--397\hfill
    DOI:~\href{https://doi.org/10.21437/Interspeech.2026-1146}{10.21437/Interspeech.2026-1146}%
  }%
  \let\@evenfoot\@oddfoot
}
\makeatother

\begin{document}

\maketitle
\thispagestyle{publication}

\begin{abstract}
Embedding enhancement improves speaker verification under acoustic mismatch without modifying a frozen backbone. Recent work has established a practical label-free setting for this task, but often adopts increasingly structured formulations. Here, the clean target is directly observed during training, making enhancement a matching problem on the unit hypersphere. We model the clean target with a von Mises--Fisher (vMF) likelihood and profile out a sample-wise concentration parameter, yielding a simple closed-form objective with adaptive weighting. Across VoxCeleb1, VoxSRC23, CN-Celeb, VOiCES, and VC-Mix, the proposed method largely preserves the baseline and gives clearer gains on challenging mismatch sets. It also remains stable under a broad single-view recipe, where a recent diffusion baseline becomes less reliable in controlled comparisons. These results suggest that effective label-free embedding enhancement in this setting does not require a highly structured formulation.
\end{abstract}

\section{Introduction}

Modern speaker verification systems map variable-length utterances to fixed-dimensional embeddings using deep neural networks~\cite{snyder2017deep,desplanques2020ecapa,liu2025masv}. These embeddings are typically trained with margin-based classification objectives~\cite{amsoftmax,aamsoftmax} or contrastive objectives~\cite{supcon,li2023discriminative}. Verification is then performed by computing the cosine similarity between embeddings. Despite this progress, acoustic mismatch remains a critical challenge. Background noise, reverberation, microphone variations, and codec effects can severely degrade performance when test conditions differ from training conditions~\cite{mclaren2016speakers,nandwana2019voices,heo2024rethinking}.

A common strategy is to build robustness directly into the pipeline through large-scale data~\cite{nagrani2017voxceleb,chung2018voxceleb2}, extensive augmentation~\cite{snyder2018x,musan,rir,specaugment}, or specialized domain adaptation~\cite{chen2020adversarial,li2020contrastiveadv,kim2025paranoise,li2025disentanglingdldae}. These approaches can be effective, but they are expensive and closely tied to the pipeline. In many applications, retraining or fine-tuning a high-capacity speaker representation extractor for each new environment is impractical.

Embedding enhancement avoids full-network retraining by applying a lightweight post-processing network to a frozen extractor. This approach is practical because it leaves the original scoring pipeline unchanged. Early work explored this direction through selective enhancement~\cite{jung2020selective} and disentangled representation learning (DRL)~\cite{drl}, though these methods often required additional supervision such as speaker labels. More recently, a generative approach based on diffusion models~\cite{nam2025seed} has established a fully label-free paradigm, demonstrating that effective enhancement can be achieved using only clean and degraded utterances.

While these recent advancements broaden the design space, they also move the field toward increasingly structured formulations. However, the fundamental appeal of embedding enhancement is precisely its lightweight nature. In this setup, the backbone is frozen, the target is directly observed during training, and inference requires only one refined embedding before cosine scoring. This makes it natural to ask how much formulation complexity is actually needed in this setting.

We show that a simple direct matching objective is already effective when aligned with the geometry of normalized embeddings. Since modern speaker embeddings are $L_2$-normalized, they lie on the unit hypersphere~\cite{garcia2011analysis,snyder2018x}. We model the clean target with a von Mises--Fisher (vMF) density and profile out a sample-wise concentration parameter, yielding a simple closed-form objective based on maximum likelihood estimation.

\begin{figure*}[t]
  \centering
  \includegraphics[width=\textwidth]{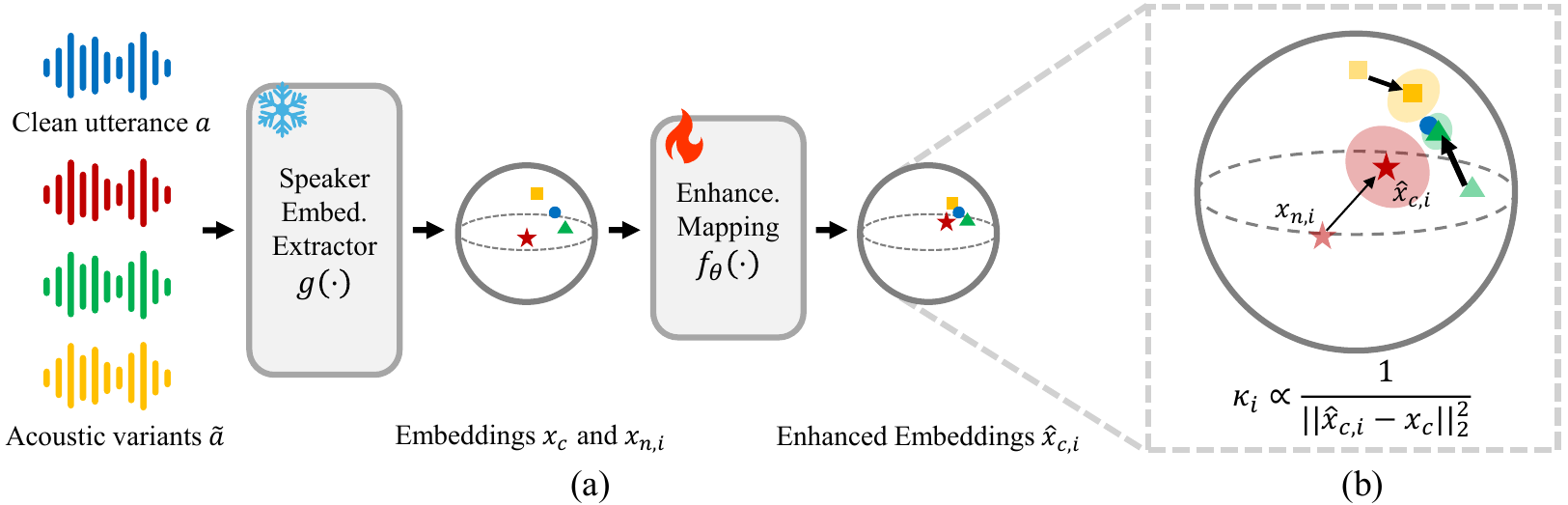}
\caption{Task overview and geometric intuition of the proposed vMF-PL objective. \textbf{(a)} An unaugmented utterance $a$ and its acoustically degraded variants $\tilde{a}$ are encoded by a frozen speaker embedding extractor $g(\cdot)$, producing a clean embedding $x_c$ and variant embeddings $x_{n,i}$. The blue marker denotes $x_c$, and the remaining colors consistently identify the corresponding acoustic variants before and after enhancement. The lightweight enhancement mapping $f_\theta(\cdot)$ moves each degraded embedding toward $x_c$ without changing the downstream cosine-scoring pipeline. \textbf{(b)} On the unit hypersphere, faded and saturated markers of the same color denote variant embeddings before and after enhancement, respectively. Arrow thickness reflects the adaptive supervision strength induced by vMF-PL, so variants closer to the clean target receive stronger supervision and are associated with higher implied concentration. Shaded regions indicate relative concentration $\kappa_i \propto 1/\lVert\hat{x}_{c,i}-x_c\rVert_2^2$.}
  \label{fig:pipeline}
\end{figure*}

The resulting loss exhibits a simple but useful property where pairs with larger residuals receive smaller updates. This is helpful when learning from heterogeneous data, reflecting the unpredictable severity of acoustic corruption found in real-world environments. Across two frozen backbones and seven evaluation sets, the proposed method largely preserves the baseline and gives clearer improvements on challenging mismatch benchmarks such as CN-Celeb, VOiCES, and VC-Mix. Figure~\ref{fig:pipeline} provides an overview of both the practical paired-target setting and the variant-wise weighting behavior induced by vMF-PL on the unit hypersphere.

In summary, our main contributions are as follows:
\begin{itemize}
  \item We revisit label-free speaker embedding enhancement in the frozen backbone setting, and show that a simple direct matching formulation is already competitive.
  \item We derive a vMF profile likelihood objective for normalized embeddings, yielding a simple closed-form loss with adaptive reweighting across training pairs.
  \item Evaluations on VoxCeleb1~\cite{nagrani2017voxceleb}, VoxSRC23~\cite{VoxSRC}, CN-Celeb~\cite{CNCeleb}, VOiCES~\cite{nandwana2019voices}, and VC-Mix~\cite{heo2024rethinking} show that the proposed method mainly preserves or improves the baseline, with clearer gains on challenging benchmarks. Under a broad single-view recipe, it also remains stable, while a recent diffusion baseline becomes unreliable in controlled comparisons.
\end{itemize}

\section{Related Works}

\subsection{Robust Speaker Verification}
Robustness to acoustic mismatch is a central issue in speaker verification~\cite{mclaren2016speakers,nandwana2019voices,sadjadi20202019,heo2024rethinking}. A standard strategy is to expose the backbone to broad acoustic variation during training~\cite{snyder2018x}, using large corpora such as VoxCeleb~\cite{nagrani2017voxceleb,chung2018voxceleb2} together with augmentations such as MUSAN~\cite{musan}, RIRs~\cite{rir}, and SpecAugment~\cite{specaugment}. These approaches directly improve the extractor and maintain strong practical baselines.

Other methods modify the network structure or training objective to handle nuisance variation more explicitly. Examples include adversarial domain adaptation to align feature distributions across mismatched conditions~\cite{chen2020adversarial,li2020contrastiveadv} and joint architectures that model speech enhancement or noise extraction together with speaker verification~\cite{kim2025paranoise}. Other recent work employs diffusion-based auto-encoding to disentangle speaker identity from frame-level speech content~\cite{li2025disentanglingdldae}. While effective, these methods remain tied to the full extraction pipeline. Adapting them to new conditions usually requires retraining or fine-tuning the backbone.

\subsection{Embedding Enhancement}
To address this limitation, embedding enhancement applies a lightweight post-processing model to a frozen extractor. Early work explored this direction through selective enhancement~\cite{jung2020selective}, which uses a gating module to suppress nuisance factors while preserving clean-condition behavior.

Subsequent work introduced disentangled representation learning (DRL) to the embedding space~\cite{drl}. This approach relies on an encoder-decoder structure and adversarial training to separate speaker identity from nuisance factors. However, it requires explicit supervision, such as speaker or session (video ID) labels, limiting its practical scalability.

More recently, SEED~\cite{nam2025seed} established a fully label-free setting by training the post-processor solely on paired clean and degraded utterances. While this avoids the need for speaker labels, it introduces a generative diffusion process directly in the embedding space. To achieve inference efficiency, it relies on single-step sampling at a fixed timestep. It also introduces the strong assumption that the diffusion trajectories of noisy and clean embeddings are nearly identical.

Our work keeps the same practical setting of a frozen backbone, a low-capacity post-processor, and training from paired clean and degraded utterances, but replaces structured generative modeling with a direct matching objective on normalized embeddings.

\section{Method}

\subsection{Task Formulation}
Modern speaker embeddings are $L_2$-normalized and evaluated using cosine similarity~\cite{garcia2011analysis,snyder2018x}. Therefore, it is natural to cast the enhancement task directly on the unit hypersphere $\mathbb{S}^{p-1}$. Let $g(\cdot)$ denote a frozen pre-trained speaker embedding extractor, and let $p$ be its embedding dimension. For an utterance $a$, let $\tilde a=\mathcal A(a)$ be a degraded counterpart obtained by injecting acoustic nuisance factors. The unaugmented $a$ defines the clean target embedding, where ``clean'' is relative to the applied degradation $\mathcal A$ rather than to studio-clean conditions. We extract the paired unit embeddings:
\begin{equation}
  x_c = \frac{g(a)}{\lVert g(a)\rVert_2}, \qquad
  x_n = \frac{g(\tilde{a})}{\lVert g(\tilde{a})\rVert_2}.
\end{equation}
The goal is to learn a deterministic enhancement mapping $f_\theta : \mathbb{S}^{p-1} \rightarrow \mathbb{S}^{p-1}$ that recovers the clean target $x_c$ from a degraded input $x_n$. By enforcing the prediction $\hat{x}_c = f_\theta(x_n)$ to closely approximate the clean target $x_c$, the mapping learns to filter out acoustic nuisance factors. At test time, the enhanced embedding $\hat{x}_c$ seamlessly integrates into the standard cosine-scoring pipeline. Since each degraded input is paired with a directly observed clean target during training, we cast enhancement as a direct matching problem on the unit hypersphere. Figure~\ref{fig:pipeline}-(a) summarizes the practical implementation of this formulation.

\subsection{vMF Profile Likelihood}
Because both $x_c$ and $\hat{x}_c$ lie on the unit hypersphere, we model the conditional distribution of the clean target using a von Mises--Fisher (vMF) density, a canonical parametric family for directional data on the unit hypersphere~\cite{banerjee2005clustering}. For the $i$-th training pair, let $\mu_i = f_\theta(x_{n,i})$. We define
\begin{equation}
  p(x_{c,i} \mid \mu_i, \kappa_i) = C_p(\kappa_i) \exp\left(\kappa_i \mu_i^\top x_{c,i}\right),
\end{equation}
where $\kappa_i \ge 0$ is a sample-wise concentration parameter and $C_p(\kappa_i)$ is the normalization constant. Let $s_i=\mu_i^\top x_{c,i}$. Profiling $\kappa_i$ gives the exact condition $A_p(\kappa_i^*)=s_i$, where $A_p(\kappa)=I_{p/2}(\kappa)/I_{p/2-1}(\kappa)$. Since $A_p^{-1}$ has no closed form, closed-form concentration approximations are standard in vMF maximum likelihood estimation~\cite{banerjee2005clustering,sra2012short}. A common approximation gives $\kappa_i^*\approx s_i(p-s_i^2)/(1-s_i^2)$, whose leading term near $s_i=1$ is $(p-1)/(2(1-s_i))$. Using the corresponding large-concentration expansion $\log C_p(\kappa_i)\approx \frac{p-1}{2}\log\kappa_i-\kappa_i$, the negative log-likelihood can be written up to additive constant as
\begin{equation}
  \ell_i(\theta, \kappa_i)
  \approx
  \kappa_i(1 - s_i) - \frac{p-1}{2}\log \kappa_i.
\end{equation}
Minimizing this objective with respect to $\kappa_i$ yields the closed-form estimate
\begin{equation}
  \kappa_i^* \approx \frac{p-1}{2(1 - s_i)}.
\end{equation}
Substituting this estimate back into the objective yields $\ell_i(\theta,\kappa_i^*) \propto \log(1-s_i)$. Since $\lVert x_{c,i}-\mu_i \rVert_2^2=2(1-s_i)$ for unit vectors, the resulting profile likelihood objective is equivalent up to additive and multiplicative constants to minimizing
\begin{equation}
  \mathcal{L}_{\text{vMF-PL}}
  =
  \frac{1}{N}\sum_{i=1}^{N}
  \log\left(\left\lVert x_{c,i} - f_\theta(x_{n,i})\right\rVert_2^2 + \epsilon\right),
  \label{eq:logmse}
\end{equation}
where $\epsilon = \eps$ ensures numerical stability. For reference, if the concentration were fixed across pairs, the vMF negative log-likelihood would reduce up to constants to standard MSE. See Section~\ref{sec:ablation_objective} for more detail.

This profile likelihood objective changes how training pairs are weighted. Its gradient with respect to the prediction is
\begin{equation}
  \frac{\partial \mathcal{L}_{\text{vMF-PL}}}{\partial \hat{x}_{c,i}}
  =
  \frac{1}{N}
  \frac{2(\hat{x}_{c,i} - x_{c,i})}
  {\lVert \hat{x}_{c,i} - x_{c,i} \rVert_2^2 + \epsilon}.
\end{equation}
The gradient magnitude is inversely scaled by the squared residual norm relative to MSE. As a result, pairs with large residuals are down-weighted, mitigating the impact of inevitable outliers found in heterogeneous real-world data. On the other hand, low-residual pairs retain relatively larger influence, helping preserve already high-quality embeddings. Figure~\ref{fig:pipeline}-(b) illustrates the sample-wise weighting behavior induced by vMF-PL on the unit hypersphere.

\section{Experiments}

\subsection{Evaluation Setup}
\textbf{Datasets and metrics.} We evaluate the proposed method using pre-trained ECAPA-TDNN~\cite{desplanques2020ecapa} and ResNet-34~\cite{chung2018voxceleb2} as frozen backbones. The VoxCeleb1 test sets (Vox1-O, Vox1-E, and Vox1-H)~\cite{nagrani2017voxceleb} serve as the standard baselines for verification performance. To assess robustness under acoustic mismatch, we also evaluate on various challenging and out of domain benchmarks, including VoxSRC23~\cite{VoxSRC}, CN-Celeb~\cite{CNCeleb}, VOiCES~\cite{nandwana2019voices}, and VC-Mix~\cite{heo2024rethinking}. We report Equal Error Rate (EER) and minimum Detection Cost Function (minDCF) with $P_{\text{target}}=0.05$.

\textbf{Comparative methods and architecture.} We compare the frozen baseline (i.e., cosine scoring directly on the unenhanced embeddings), the recent diffusion-based SEED~\cite{nam2025seed}, and the proposed vMF-PL method. All post-processing systems use the same restricted capacity architecture, 3 blocks of residual MLP with hidden width $2p$, LayerNorm, SiLU activations, and output $L_2$ normalization. This keeps model capacity fixed and focuses the comparison on the training formulation. The speaker embedding dimensions $p$ of the ECAPA-TDNN and ResNet-34 are 512 and 256, respectively. For the main comparison in Table~\ref{tab:main}, the proposed method is trained with the broad single-view recipe described in Section~\ref{sec:recipe}. In contrast, SEED is reproduced using its official implementation under its published setup, which relies on a milder multi-view training pipeline and yields its strongest performance. We adopt this asymmetric practical comparison because, in our controlled experiments, SEED did not train reliably under the broad single-view recipe used for our method. We quantify this instability explicitly in the controlled same recipe comparison of Section~\ref{sec:stability} and Table~\ref{tab:ablation_setting}. All models are trained on a single NVIDIA RTX A6000 GPU using the AdamW optimizer~\cite{adamw}. We set the learning rate to 5e-4 and apply a weight decay of 5e-5.

\subsection{Broad Single-View Training Recipe}
\label{sec:recipe}
To expose the network to diverse acoustic variations, we construct a heterogeneous training pool using the VoxCeleb2 development set (2360 hours, 1092K utterances)~\cite{chung2018voxceleb2}, LibriTTS-R (244 hours, 149.7K utterances)~\cite{librittsr}, and Libri-Light (577 hours, 2.6K utterances)~\cite{kahn2020librilight}. Since VoxCeleb2 dominates the utterance count, we randomly sample 20\% of it per epoch to maintain domain balance.

All training audio is sampled at 16 kHz and cropped to a maximum duration of 2.015 seconds. For each segment, we generate a clean-degraded pair using an overlapping augmentation pipeline:
\begin{itemize}
  \item \textbf{Environmental mismatch (50\% prob. each):} Additive noise and music from MUSAN~\cite{musan} in the range of SNR $[-20,20]$ dB, and simulated RIRs~\cite{rir}. These are applied independently and can overlap.
  \item \textbf{Telephony degradation (20\% prob.):} 16 kHz to 8 kHz down-sampling, $\mu$-law companding, and $[-5,5]$ dB random volume scaling.
\end{itemize}

Our pipeline follows the comparative method, SEED~\cite{nam2025seed}, but differs in data composition and batching. SEED utilizes only LibriTTS-R and Libri-Light, applies augmentations with milder SNR, and adopts multi-view batch construction~\cite{supcon}. Because our formulation lacks strict pipeline requirements, we adopt a memory-efficient single-view approach. Furthermore, by applying overlapping augmentations to this single view, we effectively simulate real-world degradations, e.g., background noise with reverberation. This strategy allows us to incorporate broader acoustic variations without the computational overhead of multi-view scaling.

\newcommand{\up}{\rlap{\,\raisebox{0.3ex}{\small$\uparrow$}}}
\begin{table*}[t]
  \caption{Main results on extensive benchmarks (EER (\%) / minDCF@$P_\text{target}=0.05$). Lower values are better. Bold values denote the best result for each backbone and metric. The symbol $\uparrow$ indicates performance equal to or better than the frozen baseline. All post-processing methods use the same 3-block MLP architecture. Each method is reported under the strongest stable configuration.}
  \label{tab:main}
  \centering
  \small
  \setlength{\tabcolsep}{3.5pt}
  \renewcommand{\arraystretch}{1.15}
  \resizebox{\textwidth}{!}{%
    \begin{tabular}{ll cccccc cccccccc}
      \toprule
      \multirow{2}{*}{Backbone} & \multirow{2}{*}{Method}
                                & \multicolumn{2}{c}{Vox1-O}        & \multicolumn{2}{c}{Vox1-E}        & \multicolumn{2}{c}{Vox1-H}
                                & \multicolumn{2}{c}{VoxSRC23}      & \multicolumn{2}{c}{CN-Celeb}      & \multicolumn{2}{c}{VOiCES}        & \multicolumn{2}{c}{VC-Mix} \\
      \cmidrule(lr){3-4}\cmidrule(lr){5-6}\cmidrule(lr){7-8}
      \cmidrule(lr){9-10}\cmidrule(lr){11-12}\cmidrule(lr){13-14}\cmidrule(lr){15-16}
                                &                                   & EER                       & minDCF                     & EER                       & minDCF                      & EER                       & minDCF                     & EER                       & minDCF                      & EER                       & minDCF                     & EER                       & minDCF                      & EER                       & minDCF          \\
      \midrule
      \multirow{3}{*}{ECAPA}
                                & Baseline                          & 0.91                      & 0.065                      & \textbf{1.17}             & 0.081                      & 2.39                      & \textbf{0.152}             & \textbf{5.64}             & 0.362                      & 18.28                     & \textbf{0.626}             & 6.50                      & 0.366                      & 2.96                      & 0.262          \\
                                & + SEED~\cite{nam2025seed}         & 0.94                      & 0.068                      & 1.22                      & \textbf{0.068}\up          & 2.41                      & 0.153                      & 5.96                      & 0.377                      & 17.99\up                  & 0.664                      & 6.18\up                   & \textbf{0.341}\up          & 3.02                      & 0.260\up       \\
                                & + \textbf{Ours (vMF-PL)}         & \textbf{0.88}\up          & \textbf{0.062}\up          & \textbf{1.17}\up          & 0.082                      & \textbf{2.38}\up          & \textbf{0.152}\up          & 5.65                      & \textbf{0.361}\up          & \textbf{17.89}\up         & 0.628                      & \textbf{6.17}\up          & 0.374                      & \textbf{2.82}\up          & \textbf{0.255}\up \\
      \midrule
      \multirow{3}{*}{ResNet-34}
                                & Baseline                          & \textbf{0.88}             & 0.079                      & \textbf{1.07}             & \textbf{0.076}             & 2.22                      & \textbf{0.148}             & 5.45                      & 0.335                      & 14.54                     & 0.570                      & 5.62                      & 0.300                      & 3.07                      & 0.246  \\
                                & + SEED~\cite{nam2025seed}         & 0.94                      & 0.080                      & 1.10                      & 0.079                      & 2.24                      & 0.149                      & \textbf{5.37}\up          & \textbf{0.328}\up          & 13.92\up                  & \textbf{0.569}\up          & 5.63                      & \textbf{0.294}\up          & \textbf{2.68}\up          & \textbf{0.224}\up \\
                                & + \textbf{Ours (vMF-PL)}         & \textbf{0.88}\up          & \textbf{0.077}\up          & \textbf{1.07}\up          & 0.077                      & \textbf{2.20}\up          & \textbf{0.148}\up          & 5.39\up                   & 0.335\up                   & \textbf{13.78}\up         & 0.576                      & \textbf{5.30}\up          & 0.302                      & 2.98\up                   & 0.245\up       \\
      \bottomrule
    \end{tabular}%
  }
\end{table*}

\subsection{Main Results}
Table~\ref{tab:main} reports the extensive evaluations of our method alongside the frozen backbone and SEED.  On EER, the proposed method preserves the frozen baseline on all six VoxCeleb1 entries, and matches or improves it in 13 of 14 backbone/dataset entries overall. The gains are clearer on challenging mismatch sets. With ResNet-34, EER drops from 14.54\% to 13.78\% on CN-Celeb and from 5.62\% to 5.30\% on VOiCES. With ECAPA-TDNN, EER drops from 6.50\% to 6.17\% on VOiCES and from 2.96\% to 2.82\% on VC-Mix.

Across all 28 EER/minDCF entries in Table~\ref{tab:main}, our method matches or improves the baseline on 21, compared with 12 for SEED. SEED also improves several challenging conditions, supporting the value of label-free embedding enhancement as a whole. At the same time, our method provides a more consistent balance between preserving performance on standard benchmarks and improving harder mismatch benchmarks.

\begin{table}[t]
  \caption{Controlled comparison under the same broad single-view recipe of Section~\ref{sec:recipe} with ECAPA-TDNN. EER (\%) is reported. Lower values are better.}
  \label{tab:ablation_setting}
  \centering
  \small
  \setlength{\tabcolsep}{3.5pt}
  \renewcommand{\arraystretch}{1.12}
  \resizebox{\columnwidth}{!}{%
  \begin{tabular}{lcccc}
    \toprule
    Method & Vox1-O & VoxSRC23 & VOiCES & VC-Mix \\
    \midrule
    Baseline & 0.91 & 5.64 & 6.50 & 2.96 \\
    + SEED (broad recipe) & 2.49 & 10.65 & 10.85 & 5.54 \\
    + Ours (vMF-PL) & 0.88 & 5.65 & 6.17 & 2.82 \\
    \bottomrule
  \end{tabular}%
  }
\end{table}

\subsection{Stability under a Unified Training Recipe}
\label{sec:stability}
To isolate the effect of the enhancement formulation from dataset composition and batch construction, Table~\ref{tab:ablation_setting} evaluates both methods under the same heterogeneous single-view recipe detailed in Section~\ref{sec:recipe}. Under this controlled setting, vMF-PL remains stable and reproduces the main result, whereas SEED does not train reliably. For example, its EER increases from 0.94\% to 2.49\% on Vox1-O, from 5.96\% to 10.65\% on VoxSRC23, and from 6.18\% to 10.85\% on VOiCES. This constitutes a severe degradation relative to both its published setup and the frozen baseline.

We do not interpret this as a general limitation of diffusion models. Rather, it indicates that, in the lightweight post-processing regime considered here, performance is much more sensitive to recipe design for SEED than for vMF-PL. The proposed objective is substantially more robust under broader and less curated acoustic variation.

\begin{table}[t]
  \caption{Ablation of the objective on ECAPA-TDNN. The network, data, and augmentation pipeline are fixed. EER (\%) and minDCF@$P_\text{target}=0.05$ are reported. Lower values are better.}
  \label{tab:ablation_loss}
  \centering
  \small
  \setlength{\tabcolsep}{6pt}
  \renewcommand{\arraystretch}{1.15}
  \begin{tabular}{l cc cc}
    \toprule
    \multirow{2}{*}{Loss}       & \multicolumn{2}{c}{Vox1-O} & \multicolumn{2}{c}{VoxSRC23}                                  \\
    \cmidrule(lr){2-3}\cmidrule(lr){4-5}
                                & EER                        & minDCF                       & EER           & minDCF         \\
    \midrule
    Standard MSE                & 1.05                       & 0.084                        & 6.38          & 0.413          \\
    \textbf{vMF-PL} & \textbf{0.88}              & \textbf{0.062}               & \textbf{5.65} & \textbf{0.361} \\
    \bottomrule
  \end{tabular}
\end{table}

\subsection{Ablation of the Objective}
\label{sec:ablation_objective}
Table~\ref{tab:ablation_loss} isolates the role of the objective. The network, training data, and augmentation pipeline are identical, and only the loss changes. It is worth noting that, because both the prediction and the target are $L_2$-normalized, standard MSE is not a distinct Euclidean objective in practice. Specifically, for unit vectors $x$ and $y$, $\lVert x-y\rVert_2^2 = 2 - 2x^\top y$, making MSE strictly equivalent to minimizing cosine distance up to constants.

The difference between MSE and vMF-PL therefore lies not in the underlying geometry, but in how residuals $r_i = \hat{x}_{c,i} - x_{c,i}$ are aggregated across training pairs. MSE can be viewed as vMF maximum likelihood with a fixed concentration, whereas vMF-PL profiles a sample-wise concentration and thereby induces adaptive weighting. Concretely, vMF-PL scales the MSE update by $1/(\lVert r_i\rVert_2^2 + \epsilon)$, so high-residual pairs contribute less than they would under MSE. This is especially helpful when corruption severity varies widely across training pairs, as in real-world settings. The effect is reflected in Table~\ref{tab:ablation_loss}. Replacing vMF-PL with MSE degrades Vox1-O EER from 0.88\% to 1.05\% and VoxSRC23 EER from 5.65\% to 6.38\%. The benefit therefore comes from profiled sample-wise weighting, not from additional model capacity or data.

\section{Conclusion}
We revisited label-free speaker embedding enhancement in the practical frozen-backbone setting. Because clean targets are directly observed during training, we formulate enhancement as direct matching on normalized embeddings and derive a simple vMF profile likelihood objective with adaptive weighting. Across two frozen backbones and seven benchmarks, the proposed method preserves the baseline performance, improves challenging mismatch sets, and remains more stable than both MSE and the recent diffusion baseline under the broad single-view recipe.

\section{Acknowledgments}
This work was partly supported by Institute of Information \& communications Technology Planning \& Evaluation (IITP) grant funded by the Korea government(MSIT) [NO.RS-2021-II211343, Artificial Intelligence Graduate School Program (Seoul National University)] and the National Research Foundation of Korea(NRF) grant funded by the Korea government(MSIT) [No.RS-2024-00406127, RS-2026-25477522]

\section{Generative AI Use Disclosure}
The authors used generative AI tools to improve readability and refine the language of the manuscript. After using these tools, the authors carefully reviewed, edited, and revised all content to ensure accuracy. The AI was not used to generate research ideas, analyze data, or draw scientific conclusions. The authors take full responsibility for the final content of this publication.

\bibliographystyle{IEEEtran}
\bibliography{mybib}

@article{amsoftmax,
  title={Additive margin softmax for face verification},
  author={Wang, Feng and Cheng, Jian and Liu, Weiyang and Liu, Haijun},
  journal={IEEE Signal Processing Letters},
  volume={25},
  number={7},
  pages={926--930},
  year={2018},
  publisher={IEEE}
}

@inproceedings{aamsoftmax,
  title={Arcface: Additive angular margin loss for deep face recognition},
  author={Deng, Jiankang and Guo, Jia and Xue, Niannan and Zafeiriou, Stefanos},
  booktitle={Proceedings of the IEEE/CVF conference on computer vision and pattern recognition},
  pages={4690--4699},
  year={2019}
}

@article{supcon,
  title={Supervised contrastive learning},
  author={Khosla, Prannay and Teterwak, Piotr and Wang, Chen and Sarna, Aaron and Tian, Yonglong and Isola, Phillip and Maschinot, Aaron and Liu, Ce and Krishnan, Dilip},
  journal={Advances in neural information processing systems},
  volume={33},
  pages={18661--18673},
  year={2020}
}

@article{musan,
  title={Musan: A music, speech, and noise corpus},
  author={Snyder, David and Chen, Guoguo and Povey, Daniel},
  journal={arXiv preprint arXiv:1510.08484},
  year={2015}
}

@inproceedings{specaugment,
  title={SpecAugment: A Simple Data Augmentation Method for Automatic Speech Recognition},
  author={Park, Daniel S and Chan, William and Zhang, Yu and Chiu, Chung-Cheng and Zoph, Barret and Cubuk, Ekin D and Le, Quoc V},
  booktitle={Proc. Interspeech 2019},
  pages={2613--2617},
  year={2019}
}

@inproceedings{librittsr,
  title={LibriTTS-R: A Restored Multi-Speaker Text-to-Speech Corpus},
  author={Koizumi, Yuma and Zen, Heiga and Karita, Shigeki and Ding, Yifan and Yatabe, Kohei and Morioka, Nobuyuki and Bacchiani, Michiel and Zhang, Yu and Han, Wei and Bapna, Ankur},
  booktitle={Proc. Interspeech 2023},
  pages={5496--5500},
  year={2023}
}

@inproceedings{rir,
  title={A study on data augmentation of reverberant speech for robust speech recognition},
  author={Ko, Tom and Peddinti, Vijayaditya and Povey, Daniel and Seltzer, Michael L and Khudanpur, Sanjeev},
  booktitle={2017 IEEE international conference on acoustics, speech and signal processing (ICASSP)},
  pages={5220--5224},
  year={2017},
  organization={IEEE}
}

@inproceedings{drl,
  title={Disentangled Representation Learning for Environment-agnostic Speaker Recognition},
  author={Nam, KiHyun and Heo, Hee-Soo and Jung, Jee-weon and Chung, Joonson},
  booktitle={Proc. Interspeech 2024},
  pages={2130--2134},
  year={2024}
}

@inproceedings{liu2025masv,
  title={MASV: Speaker Verification with Global and Local Context Mamba},
  author={Liu, Yang and Wan, Li and Huang, Yiteng and Sun, Ming and Mei, Xinhao and Liu, Xubo and Shi, Yangyang and Metze, Florian},
  booktitle={Proc. Interspeech 2025},
  pages={3703--3707},
  year={2025}
}

@inproceedings{snyder2017deep,
  title={Deep neural network embeddings for text-independent speaker verification.},
  author={Snyder, David and Garcia-Romero, Daniel and Povey, Daniel and Khudanpur, Sanjeev},
  booktitle={Interspeech},
  volume={2017},
  pages={999--1003},
  year={2017}
}

@inproceedings{desplanques2020ecapa,
  title={ECAPA-TDNN: Emphasized Channel Attention, Propagation and Aggregation in TDNN Based Speaker Verification},
  author={Desplanques, Brecht and Thienpondt, Jenthe and Demuynck, Kris},
  booktitle={Proc. Interspeech 2020},
  pages={3830--3834},
  year={2020}
}

@inproceedings{li2023discriminative,
  title={Discriminative speaker representation via contrastive learning with class-aware attention in angular space},
  author={Li, Zhe and Mak, Man-Wai and Meng, Helen Mei-Ling},
  booktitle={ICASSP 2023-2023 IEEE International Conference on Acoustics, Speech and Signal Processing (ICASSP)},
  pages={1--5},
  year={2023},
  organization={IEEE}
}

@inproceedings{jung2020selective,
  title={Selective Deep Speaker Embedding Enhancement for Speaker Verification},
  author={Jung, Jee-Weon and Kim, Ju-Ho and Shim, Hye-Jin and Kim, Seung-bin and Yu, Ha-Jin},
  booktitle={Proc. Odyssey 2020},
  pages={171--178},
  year={2020}
}

@inproceedings{kim2025paranoise,
  title={ParaNoise-SV: Integrated Approach for Noise-Robust Speaker Verification with Parallel Joint Learning of Speech Enhancement and Noise Extraction},
  author={Kim, Minu and Jang, Kangwook and Kim, Hoirin},
  booktitle={Proc. Interspeech 2025},
  pages={1103--1107},
  year={2025}
}

@inproceedings{li2025disentanglingdldae,
  title={Disentangling speaker and content in pre-trained speech models with latent diffusion for robust speaker verification},
  author={Li, Zhe and Mak, Man-Wai and Chien, Jen-Tzung and Pilanci, Mert and Jin, Zezhong and Meng, Helen},
  booktitle={Proc. Interspeech},
  pages={1108--1112},
  year={2025}
}

@inproceedings{nam2025seed,
  title={SEED: Speaker Embedding Enhancement Diffusion Model},
  author={Nam, Kihyun and Heo, Jungwoo and Jung, Jee-weon and Park, Gangin and Jung, Chaeyoung and Yu, Ha-Jin and Chung, Joon Son},
  booktitle={Proc. Interspeech 2025},
  pages={3718--3722},
  year={2025}
}

@inproceedings{nandwana2019voices,
  title={The VOiCES from a Distance Challenge 2019},
  author={Nandwana, Mahesh Kumar and Hout, Julien van and Richey, Colleen and McLaren, Mitchell and Barrios, Maria A and Lawson, Aaron},
  booktitle={Proc. Interspeech 2019},
  pages={2438--2442},
  year={2019}
}

@inproceedings{sadjadi20202019,
  title={The 2019 NIST Speaker Recognition Evaluation CTS Challenge},
  author={Sadjadi, Seyed Omid and Greenberg, Craig and Singer, Elliot and Reynolds, Douglas and Mason, Lisa and Hernandez-Cordero, Jaime},
  booktitle={Proc. Odyssey 2020},
  pages={266--272},
  year={2020}
}

@inproceedings{mclaren2016speakers,
  title={The Speakers in the Wild (SITW) Speaker Recognition Database},
  author={McLaren, Mitchell and Ferrer, Luciana and Castan, Diego and Lawson, Aaron},
  booktitle={Proc. Interspeech 2016},
  pages={818--822},
  year={2016}
}

@inproceedings{heo2024rethinking,
  title={Rethinking session variability: Leveraging session embeddings for session robustness in speaker verification},
  author={Heo, Hee-Soo and Nam, KiHyun and Lee, Bong-Jin and Kwon, Youngki and Lee, Minjae and Kim, You Jin and Chung, Joon Son},
  booktitle={ICASSP 2024-2024 IEEE International Conference on Acoustics, Speech and Signal Processing (ICASSP)},
  pages={12321--12325},
  year={2024},
  organization={IEEE}
}

@inproceedings{snyder2018x,
  title={X-vectors: Robust dnn embeddings for speaker recognition},
  author={Snyder, David and Garcia-Romero, Daniel and Sell, Gregory and Povey, Daniel and Khudanpur, Sanjeev},
  booktitle={2018 IEEE international conference on acoustics, speech and signal processing (ICASSP)},
  pages={5329--5333},
  year={2018},
  organization={IEEE}
}

@inproceedings{garcia2011analysis,
  title={Analysis of i-vector length normalization in speaker recognition systems.},
  author={Garcia-Romero, Daniel and Espy-Wilson, Carol Y},
  booktitle={Interspeech},
  volume={2011},
  pages={249--252},
  year={2011}
}

@article{li2020contrastiveadv,
  title={Contrastive adversarial domain adaptation networks for speaker recognition},
  author={Li, Longxin and Mak, Man-Wai and Chien, Jen-Tzung},
  journal={IEEE Transactions on Neural Networks and Learning Systems},
  volume={33},
  number={5},
  pages={2236--2245},
  year={2020},
  publisher={IEEE}
}

@article{chen2020adversarial,
  title={Adversarial domain adaptation for speaker verification using partially shared network},
  author={Chen, Zhengyang},
  journal={Interspeech 2020},
  year={2020}
}

@article{banerjee2005clustering,
  title={Clustering on the Unit Hypersphere using von Mises-Fisher Distributions.},
  author={Banerjee, Arindam and Dhillon, Inderjit S and Ghosh, Joydeep and Sra, Suvrit and Ridgeway, Greg},
  journal={Journal of Machine Learning Research},
  volume={6},
  number={9},
  year={2005}
}

@inproceedings{nagrani2017voxceleb,
  author    = {Nagrani, Arsha and Chung, Joon Son and Zisserman, Andrew},
  title     = {VoxCeleb: a Large-Scale Speaker Identification Dataset},
  booktitle = {Proc. Interspeech},
  year      = {2017},
  pages     = {2616--2620}
}

@inproceedings{chung2018voxceleb2,
  author    = {Chung, Joon Son and Nagrani, Arsha and Zisserman, Andrew},
  title     = {VoxCeleb2: Deep Speaker Recognition},
  booktitle = {Proc. Interspeech},
  year      = {2018},
  pages     = {1086--1090}
}

@inproceedings{CNCeleb,
 title={Cn-celeb: a challenging chinese speaker recognition dataset}, 
 author={Fan, Yue and Kang, JW and Li, LT and Li, KC and Chen, HL and Cheng, ST and Zhang, PY and Zhou, ZY and Cai, YQ and Wang, Dong},
 booktitle={ICASSP 2020-2020 IEEE International Conference on Acoustics, Speech and Signal Processing (ICASSP)},
 pages={7604--7608}, year={2020}, organization={IEEE} }

@article{VoxSRC,
  title={The VoxCeleb Speaker Recognition Challenge: A Retrospective},
  author={Huh, Jaesung and Chung, Joon Son and Nagrani, Arsha and Brown, Andrew and Jung, Jee-weon and Garcia-Romero, Daniel and Zisserman, Andrew},
  journal={IEEE/ACM Transactions on Audio, Speech, and Language Processing},
  year={2024},
  publisher={IEEE}
}

@article{adamw,
  title={Decoupled weight decay regularization},
  author={Loshchilov, Ilya and Hutter, Frank},
  journal={arXiv preprint arXiv:1711.05101},
  year={2017}
}

@inproceedings{kahn2020librilight,
  title={Libri-light: A benchmark for asr with limited or no supervision},
  author={Kahn, Jacob and Riviere, Morgane and Zheng, Weiyi and Kharitonov, Evgeny and Xu, Qiantong and Mazar{\'e}, Pierre-Emmanuel and Karadayi, Julien and Liptchinsky, Vitaliy and Collobert, Ronan and Fuegen, Christian and others},
  booktitle={ICASSP 2020-2020 IEEE International Conference on Acoustics, Speech and Signal Processing (ICASSP)},
  pages={7669--7673},
  year={2020},
  organization={IEEE}
}

@article{sra2012short,
  title={A short note on parameter approximation for von {Mises--Fisher} distributions:
  and a fast implementation of {$I_s(x)$}},
  author={Sra, Suvrit},
  journal={Computational Statistics},
  volume={27},
  number={1},
  pages={177--190},
  year={2012},
  doi={10.1007/s00180-011-0232-x}
}

\end{document}